\documentclass[iop]{emulateapj}

\usepackage{natbib,graphicx,amsmath}

\usepackage[usenames]{color}

\shorttitle{SPECIAL RELATIVISTIC MHD WITH GRAVITATION}
\shortauthors{NOH, HWANG \& BUCHER}

\newcommand{\bea}{\begin{eqnarray}}
\newcommand{\eea}{\end{eqnarray}}

\begin{document}

\title{Special Relativistic Magnetohydrodynamics with Gravitation}
\author{Hyerim Noh${}^{1}$,
        Jai-chan Hwang${}^{2}$,
        Martin Bucher${}^{3,4}$}
\address{${}^{1}$Center for Large Telescope,
         Korea Astronomy and Space Science Institute, Daejon, Korea \\
         ${}^{2}$Department of Astronomy and Atmospheric Sciences, Kyungpook National University, Daegu, Korea \\
         ${}^{3}$APC, AstroParticule et Cosmologie, Universit\'e Paris Diderot,
         CNRS/IN2P3, CEA/lrfu, Observatoire de Paris, Sorbonne Paris Cit\'e\\
         10, rue Alice Domon et L\'eonie Duquet, 75205 Paris Cedex 13, France\\
         ${}^{4}$Astrophysics and Cosmology Research Unit, University of KwaZulu-Natal, Durban, 4041, South Africa
         }


\begin{abstract}

We present a fully nonlinear and exact perturbation formulation of Einstein's gravity with a
general fluid and the ideal magnetohydrodynamics (MHD) without imposing the slicing (temporal
gauge) condition.  Using this formulation, we derive equations of special relativistic (SR) MHD
in the presence of weak gravitation.  The equations are consistently derived in the limit of weak
gravity and action-at-a-distance in the maximal slicing. We show that in this approximation
the relativistic nature of gravity does not affect the SR MHD dynamics, but SR effects manifest themselves
in the metric, and thus gravitational lensing. Neglecting these SR effects may lead to an
overestimation of lensing masses.

\end{abstract}

\noindent 

\maketitle

\section{Introduction}

Magnetic fields are ubiquitous in celestial objects and in the universe as a whole.
Magnetohydrodynamics (MHD) is often a useful approximation for treating fluid and gas
coupled to the electromagnetic field.
Relativistic processes play a crucial role in many astrophysical phenomena.
Relativistic MHD is required to understand the physical processes in accretion disks,
magnetospheres, the plasma winds and astrophysical jets near compact objects
(e.g., neutron stars and black holes), and active galactic nuclei where the relativistic effects
(of gravity, the gas velocity, the field strength, etc.) are significant.
In none of these situation can the astrophysical processes be treated ignoring gravity.

Special relativistic (SR) MHD with non-relativistic gravity has been studied in the
literature as well as fully generally relativistic (GR) MHD,
which is implemented in numerical relativity simulations.
(For textbook treatments, see Bona et al. 2009; Baumgarte \& Shapiro 2010;
Gourgoulhon 2012; Shibata 2016).
Full blown simulations using numerical relativity are ultimately the most reliable
technique, but are computationally complicated, time consuming, and expensive.
Moreover, the results depend on the gauge choice, gauge (i.e., coordinate condition),
often making it difficult to extract the proper physical interpretation.

Here we present two formulations of GR MHD.  The first is the exact and fully nonlinear GR MHD
perturbation equations without imposing a particular
slicing, or temporal gauge condition. This is an
extension of fully nonlinear and exact perturbation formulation in the
cosmology context (Hwang
\& Noh 2013; Noh 2014, Hwang, Noh \& Park 2016) now including ideal MHD in Minkowski
background. The other formulation is the SR MHD with weak gravitation as a consistent limit of the
fully nonlinear and exact GR-MHD formulation [for hydrodynamic case, see Hwang \& Noh
(2016)].

The equations of SR MHD with weak gravity and the fully nonlinear and exact perturbation
of GR MHD are summarized in Sec.\ \ref{sec:SRG} and in the Appendix, respectively. The equations are
derived in Secs.\ \ref{sec:SR-MHD-derivation} and \ref{sec:FNLE-MHD-derivation},
respectively. In our approximation the relativistic nature of gravity does not affect the SR MHD
dynamics, see Sec.\ \ref{sec:SRG-role-of-gravity}, but SR effects appear in the metric,
thus affecting gravitational lensing, see Sec.\ \ref{sec:SRG-lensing}.

\section{Special relativistic MHD with gravitation}
                                         \label{sec:SRG}

Our metric convention is
\bea
   & &
       ds^2 = - \left( 1 + 2 \alpha \right) c^2 d t^2
       - 2 \chi_i c dt d x^i
   \nonumber \\
   & & \qquad
       + \left( 1 + 2 \varphi \right) \delta_{ij} d x^i d x^j.
   \label{metric}
\eea
The index of $\chi_i$ is raised and lowered
using
$\delta_{ij}$ as the metric. Our spatial
gauge conditions together with {\it neglecting} the (transverse-traceless) gravitational
wave mode allow us to write the spatial part of the metric as above (Hwang \& Noh 2013). As
we have not assumed any condition on $\alpha$, $\varphi$ and $\chi_i,$ our formulation is
valid to fully nonlinear order and exact. An extension to include the transverse, traceless mode
without
imposing
the spatial gauge conditions is presented in Gong et al.~(2017).

Keeping the $\chi_i$ term is important to achieve a consistent derivation of our result
in the weak gravity limit. We
{\it take} the maximal slicing condition by setting the trace of the extrinsic curvature to
be zero; in our notation $\kappa \equiv 0$ where $\kappa$ is defined in Eq.\ (\ref{eq1});
$\kappa $ is the same as the expansion scalar in the normal frame with a minus sign, thus often
termed as the uniform-expansion gauge in cosmology.

We {\it assume} the following weak gravity and action-at-a-distance conditions\footnote{We
reverse the signs of $\Phi$ and $\Psi$ relative to the convention in Hwang \& Noh (2016).}
\bea
   & & \alpha \equiv {\Phi \over c^2} \ll 1, \quad
       \varphi \equiv - {\Psi \over c^2} \ll 1, \quad
       \gamma^2 {t_\ell^2 \over t_g^2} \ll 1,
   \label{WG-conditions}
\eea
where $t_g \sim 1/\sqrt{G \varrho}$ and $t_\ell \sim \ell/c \sim 2 \pi /(kc)$ are
gravitational timescale and the light
propagation
timescale of a characteristic length
scale $\ell$, respectively; $k$ is the wave number with $\Delta = - k^2$.

\subsection{Results}

The equations of motion of SR MHD in the presence of weak gravity are
\bea
   & & {\partial \over \partial t}
       \left(
       \begin{array}{c}
           D   \\
           E   \\
           m^i \\
           B^i
       \end{array}
       \right)
       + \nabla_j
       \left(
       \begin{array}{c}
           D v^j  \\
           m^j c^2   \\
           m^{ij} \\
           v^j B^i - v^i B^j
       \end{array}
       \right)
   \nonumber \\
   & & \qquad
       =
       \left(
       \begin{array}{c}
           0   \\
           - \overline \varrho \left( 2 \Phi - \Psi \right)_{,i}v^i   \\
           - \overline \varrho \Phi^{,i} \\
           0
       \end{array}
       \right),
   \label{SR-MHD-eqs}
\eea
and
\bea
   & & B^i_{\;\;,i} = 0.
   \label{Maxwell-eq3}
\eea
These are the mass, energy, and momentum conservation equations, and the two Maxwell equations,
respectively.

The notation is as follows:
\bea
   & &
       D \equiv \overline \varrho \gamma, \quad
       \varrho \equiv \overline \varrho \left(
       1 + {\Pi \over c^2} \right),
   \nonumber \\
   & & E/c^2 \equiv \left( \varrho + {p \over c^2} \right) \gamma^2
       - {p \over c^2}
       + {1 \over c^4} \Pi_{ij} v^i v^j
   \nonumber \\
   & & \qquad
       + {1 \over 8 \pi} {1 \over c^2}
       \left[ B^2 \left( 1 + {v^2 \over c^2} \right)
       - {1 \over c^2} \left( B^i v_i \right)^2 \right],
   \nonumber \\
   & & m^i \equiv \left( \varrho + {p \over c^2} \right)
       \gamma^2 v^i
   \nonumber \\
   & & \qquad
       + {1 \over c^2} \left[ \Pi^i_j v^j
       + {1 \over 4 \pi} \left( B^2 v^i
       - B^i B^j v_j \right) \right],
   \nonumber \\
   & & m^{ij} \equiv \left( \varrho + {p \over c^2} \right) \gamma^2
       v^i v^j
       + p \delta^{ij}
       + \Pi^{ij}
   \nonumber \\
   & & \qquad
       + {1 \over 4 \pi} \bigg\{
       {1 \over \gamma^2} \left( {1 \over 2} B^2 \delta^{ij}
       - B^i B^j \right)
       + {1 \over c^2} \bigg[ B^2 v^i v^j
   \nonumber \\
   & & \qquad
       +
       {1 \over 2} \left( B^k v_k \right)^2 \delta^{ij}
       - \left( B^j v^i + B^i v^j \right) B^k v_k
       \bigg] \bigg\}.
   \label{D-def}
\eea
$\varrho$ denotes the density, $\overline \varrho$ the mass density,
$\overline \varrho \Pi$ the internal energy, $p$ the pressure, $\Pi_{ij}$
the anisotropic stress, $v^i$ the velocity, $\gamma$ the Lorentz factor
\bea
   & & \gamma
       = {1 \over \sqrt{ 1 - {v^2 \over c^2}}},
\eea
and $B_i$ the magnetic field. For a more general form of $\gamma$
with strong gravity, see Eq.\ (\ref{Lorentz-factor}). We have
\bea
   & & \Pi^i_i = \Pi_{ij} {v^i v^j \over c^2}.
\eea
Contributions from gravity appear in the right-hand side of Eq.\ (\ref{SR-MHD-eqs}).
All spatial indices in this section are raised and lowered using $\delta_{ij}$ as the metric.
Following the notation typically used in the ADM formulation of GR,
we define $E$ as the ADM energy density, $J_i = c m_i$ the ADM flux vector,
$S_{ij} = m_{ij}$ the ADM stress where $S \equiv S^i_i = m^i_i$ is its isotropic part.
Indices of the ADM fluid variables are raised and lowered by the ADM metric
$h_{ij}$ and this is the same as $\delta_{ij}$ in our approximation.
Using
\bea
   & & {\bf E}
       = - {1 \over c}
       {\bf v} \times {\bf B}
\eea
in the ideal MHD approximation, the electromagnetic (EM) parts of above quantities can be written as
\bea
   & & E_{\rm MHD} = {\cal S}_{\rm MHD}
       = {1 \over 8 \pi} \left( E^2 + B^2 \right),
   \nonumber \\
   & & m^i_{\rm MHD}
       = {1 \over 4 \pi c}
       \left( {\bf E} \times {\bf B} \right)^i,
   \nonumber \\
   & & m^{ij}_{\rm MHD}
       = {1 \over 4 \pi} \left[ - E^i E^j - B^i B^j
       + {1 \over 2} \left( E^2 + B^2 \right) \delta^{ij} \right].
\eea

The left-hand side of Eq.\ (\ref{SR-MHD-eqs}) is exactly the same as
for
SR MHD
(Mignone et al. 2007) with an anisotropic stress, and the right-hand side is
the source term due to gravity. The two gravitational potentials in the metric
satisfy the two Poisson-like equations
\bea
   & & \Delta \Phi
       = 4 \pi G \left( \varrho + {3 p \over c^2}
       + {2 \over c^2} {\cal S} \right)
       = 4 \pi G {E + S \over c^2},
   \label{eq4-SRG} \\
   & & \Delta \Psi
       = 4 \pi G \left( \varrho
       + {1 \over c^2} {\cal S} \right)
       = 4 \pi G {E \over c^2},
   \label{eq2-SRG}
\eea
with
\bea
   & & {\cal S} \equiv
       \left( \varrho + {p \over c^2} \right) \gamma^2 {v^2}
       + \Pi_{ij} {v^i v^j \over c^2}
   \nonumber \\
   & & \qquad
       + {1 \over 8 \pi}
       \left[ B^2 + {1 \over c^2}
       \left( {\bf v} \times {\bf B} \right)^2 \right],
   \label{S-SRG} \\
   & & E = \varrho c^2 + {\cal S}, \quad
       S = 3 p + {\cal S}.
\eea
The metric component $\chi_i$ is determined by
\bea
   & & \chi_i = - {4 \pi G \over c^3} \Delta^{-1}
       \left( 4 \delta_i^j
       - \Delta^{-1} \nabla_i \nabla^j \right) m_j.
   \label{eq3-SRG}
\eea
Equations (\ref{eq4-SRG}) and (\ref{eq2-SRG}) show that the weak gravity
conditions imply the action-at-a-distance condition in Eq.\ (\ref{WG-conditions}).

Equations (\ref{SR-MHD-eqs})-(\ref{eq3-SRG}) constitute a complete set.
The pressure and anisotropic stress should be specified by equations of state, and we
do not consider additional presence of heat flux.
All the above equations are consistently derived in
Sec.\ \ref{sec:SR-MHD-derivation} from a fully GR MHD formulation
derived and presented in Sec.\ \ref{sec:FNLE-MHD-derivation} and in the Appendix, respectively.

\subsection{Role of relativistic gravity on dynamics}
                                         \label{sec:SRG-role-of-gravity}

In the derivation of the first three conservation equations in Eq.\ (\ref{SR-MHD-eqs}),
we have strictly imposed the conditions in Eq.\ (\ref{WG-conditions}).
All terms in Eqs.\ (\ref{D-def}), (\ref{eq4-SRG}), (\ref{eq2-SRG}), and (\ref{eq3-SRG})
are of the same order as in the fully SR situation with
\bea
   & & 1
       \sim {\varrho \over \overline \varrho}
       \sim {\Pi \over c^2}
       \sim {v^2 \over c^2}
       \sim {p \over \overline \varrho c^2}
       \sim {\Pi_{ij} \over \overline \varrho c^2}
       \sim {B^2 \over \overline \varrho c^2}.
   \label{relativistic-order}
\eea
Applying the weak gravity condition, we find that Poisson's equation simply becomes
\bea
   & & \Delta \Phi
       = 4 \pi G \overline \varrho,
\eea
with $\Psi = \Phi$. Thus the gravity part in Eq.\ (\ref{SR-MHD-eqs}) becomes
\bea
   & & \left(
       \begin{array}{c}
           0   \\
           - \overline \varrho \Phi_{,i}v^i   \\
           - \overline \varrho \Phi^{,i} \\
           0
       \end{array}
       \right).
\eea
Therefore, in the framework of our approximation, the relativistic nature of
gravity does not alter the dynamics of fluid and fields.

For a static equilibrium situation, we have $v^i = 0$, and the momentum conservation equation gives
\bea
   & & \nabla_j \left[
       \left( p + {1 \over 8 \pi} B^2 \right) \delta^{ij}
       + \Pi^{ij} - {1 \over 4 \pi} B^i B^j \right]
       = - \overline \varrho \Phi^{,i}.
\eea
For the gravitational potential in the above equation,
we have $\Delta \Phi = 4 \pi G \overline \varrho$. However, for the metric we have
\bea
   & & \Delta \Phi = 4 \pi G \left( \varrho
       + {3 p \over c^2} + {1 \over 4 \pi} B^2 \right),
   \\
   & & \Delta \Psi = 4 \pi G \left( \varrho
       + {1 \over 8 \pi} B^2 \right),
\eea
and $\chi_i = 0$. The metric is curved by the pressure and magnetic field contributions
as well as the mass density, and these extra contributions
alter
the gravitational lensing predictions.

\subsection{Impact of special relativity on gravitational lensing}
                                         \label{sec:SRG-lensing}

Equations (\ref{eq4-SRG})-(\ref{eq3-SRG}) determine the spacetime metric
calculated assuming weak gravity and taking into account SR effects,
which are included in our approximation. Although $\chi_i$ is non-vanishing in the maximal slicing,
we can show that in the weak gravity approximation, the null geodesic equation simply becomes
\bea
   & & {d^2 x^i \over d t^2}
       = - \left( \Phi + \Psi \right)^{,i},
\eea
and thus is the same as in the zero-shear gauge, taking $\chi \equiv 0$ as the slicing condition.
The null geodesic equation to 1PN order can be found in Sec.\ 5 of Hwang et al (2008) using a notation
following Noh \& Hwang (2013).

We note that the special relativistic effects of velocity, internal energy, pressure, anisotropic
stress, and the magnetic field cause the two potentials $\Phi$ and $\Psi$
to differ from each other.
This might cause the gravitational lensing to differ from the conventional result, which assumes
$\Psi = \Phi$.
In addition to
this asymmetric effect (often known as a gravitational slip of
the potentials), in the presence of this SR effect, instead of
$\Delta (\Phi + \Psi) = 8 \pi G \overline \varrho$, we have
\bea
   & & \Delta \left( \Phi + \Psi \right)
       = 4 \pi G \bigg\{ 2 \overline \varrho \bigg(
       1 + {\Pi \over c^2} \bigg)
   \nonumber \\
   & & \qquad
       + {3 \over c^2} \bigg[ p
       + \left( \varrho + {p \over c^2} \right) \gamma^2 {v^2}
       + \Pi_{ij} {v^i v^j \over c^2}
   \nonumber \\
   & & \qquad
       + {1 \over 8 \pi}
       \bigg( B^2 + {1 \over c^2}
       \left( {\bf v} \times {\bf B} \right)^2 \bigg) \bigg] \bigg\}.
   \label{lensing-potential}
\eea
The gravitational potential $2 \Phi$ in
the commonly used
gravitational lensing
formulae in Einstein's gravity should be replaced with $\Phi + \Psi$.
For positive pressure and anisotropic stress, all the SR terms
leads to an
overestimation of the mass. In a homogeneous background medium
(as in Friedmann cosmology)
with linear perturbations, only the density and pressure terms contribute
to the lensing. The other terms are nonlinear perturbations.

For a negative pressure with an equation of state approximating dark energy
with $p_{\rm DE} \simeq - \varrho_{\rm DE} c^2$,
we have $\Delta ( \Phi + \Psi ) \simeq - 4 \pi G \varrho_{\rm DE}$.
If this component is clustered,
de-lensing by dispersing the light may result. In the presence of ordinary matter and
a dark energy component, we have
$\Delta ( \Phi + \Psi ) \simeq 4 \pi G ( 2 \varrho_{\rm matter} - \varrho_{\rm DE})$
where
$\varrho_{\rm DE} \simeq \Lambda c^2/(8 \pi G)$ and $\Lambda$ is the cosmological constant.

The presence of pressure term
in the relativistic Poisson's equation in (\ref{eq4-SRG})
has often been
noted on several occasions
in the literature
(Tolman 1930; Whittaker 1935; McCrea 1951; Harrison 1965).
The pressure term, however, vanishes in the zero-shear gauge, and this
contradicts with the well known Tolman-Oppenheimer-Volkoff equation
for a spherically symmetric static solution
(Tolman 1939; Oppenheimer \& Volkoff 1939).
In Hwang \& Noh (2016) we
argued that
when relativistic pressure is present,
the zero-shear gauge is not
a suitable gauge choice because it leads to an inconsistent result.

\subsection{Non-relativistic MHD limit}
                                         \label{sec:SRG-NR-limit}

As the non-relativistic limit, we take $c \rightarrow \infty$.
To get the energy conservation equation properly, we need to
consider next order in $c^{-2}$; this is because our $E$ contains
the rest mass energy density which satisfies the continuity
equation separately.
In other words, in the $c \rightarrow \infty$ limit,
$(E - D c^2)^{\displaystyle\cdot} + \nabla_j (m^j - D v^j) c^2
= - \overline \varrho \Phi_{,i}v^i$ gives
\bea
   & & \left( {1 \over 2} \overline \varrho v^2
       + \overline \varrho \Pi + {1 \over 8 \pi} B^2 \right)^{\displaystyle{\cdot}}
       + \nabla_i \bigg\{
       \left( {1 \over 2} \overline \varrho v^2
       + \overline \varrho \Pi + p \right) v^i
   \nonumber \\
   & & \qquad
       + \Pi^i_j v^j
       - {1 \over 4 \pi} \left[ \left( {\bf v} \times {\bf B} \right)
       \times {\bf B} \right]^i \bigg\}
       = - \overline \varrho {\bf v} \cdot \nabla \Phi.
   \label{E-conservation-NR}
\eea
The complete set of equations is
\bea
   & & \dot {\overline \varrho}
       + \nabla \cdot \left( \overline \varrho {\bf v} \right)
       = 0,
   \label{continuity-NR} \\
   & & \dot \Pi
       + {\bf v} \cdot \nabla \Pi
       + {1 \over \overline \varrho}
       \left( p \nabla \cdot {\bf v} + v_{i,j} \Pi^{ij} \right)
       = 0,
   \\
   & & \overline \varrho \left( \dot {\bf v}
       + {\bf v} \cdot \nabla {\bf v} \right)
       = - \overline \varrho \nabla \Phi
       - \nabla p - \nabla_j \Pi^{ij}
   \nonumber \\
   & & \qquad
       + {1 \over 4 \pi}
       \left( \nabla \times {\bf B} \right) \times {\bf B},
   \label{mom-conservation-NR} \\
   & & \dot {\bf B} = \nabla \times
       \left( {\bf v} \times {\bf B} \right), \quad
       \nabla \cdot {\bf B} = 0,
   \label{Faraday-NR} \\
   & & \Delta \Phi = 4 \pi G \overline \varrho,
\eea
and we have
\bea
   & & {\bf E} = - {1 \over c} {\bf v} \times {\bf B}.
\eea

Combining Eqs.\ (\ref{continuity-NR}) and (\ref{mom-conservation-NR}) we have
\bea
   & & \left( \overline \varrho v^i \right)^{\displaystyle{\cdot}}
       + \nabla_j \bigg[ \overline \varrho v^i v^j
       + \left( p + {1 \over 8 \pi} B^2 \right) \delta^{ij}
       + \Pi^{ij} - {1 \over 4 \pi} B^i B^j \bigg]
   \nonumber \\
   & & \qquad
       = - \overline \varrho \Phi^{,i},
   \label{M-conservation-NR}
\eea
where
the contributions of magnetic field are interpreted as magnetic
pressure and magnetic tension force densities, respectively. These
differ from the contribution to the pressure and anisotropic stress
appearing
in the energy-momentum tensor; in the non-relativistic limit,
Eq.\ (\ref{fluid-MHD}) implies that
\bea
   & & \mu^{\rm MHD}
       = 3 p^{\rm MHD}
       = {1 \over 8 \pi} B^2,
   \nonumber \\
   & & \Pi^{\rm MHD}_{ij}
       = - {1 \over 4 \pi}
       \left( B_i B_j - {1 \over 3} \delta_{ij} B^2 \right).
   \label{fluid-MHD-NR}
\eea
By replacing $\mu \rightarrow \mu + \mu^{\rm MHD}$, etc., in the
hydrodynamic equations, we can derive the MHD equations.

The above equations can be combined to give
\bea
   & & {\partial \over \partial t}
       \left(
       \begin{array}{c}
           \overline \varrho  \\
           \overline E  \\
           \overline \varrho v^i \\
           B^i
       \end{array}
       \right)
       + \nabla_j
       \left(
       \begin{array}{c}
           \overline \varrho v^j  \\
           \overline m^j c^2  \\
           m^{ij} \\
           v^j B^i - v^i B^j
       \end{array}
       \right)
       =
       \left(
       \begin{array}{c}
           0   \\
           - \overline \varrho \Phi_{,i}v^i   \\
           - \overline \varrho \Phi^{,i} \\
           0
       \end{array}
       \right),
   \label{SR-MHD-eqs-NR}
\eea
with
\bea
   & & \overline E
       \equiv E - D c^2
       = {1 \over 2} \overline \varrho v^2
       + \overline \varrho \Pi + {B^2 \over 8 \pi},
   \nonumber \\
   & & \overline m^i c^2
       \equiv \left( m^i - D v^i \right) c^2
       = \left( {1 \over 2} \overline \varrho v^2
       + \overline \varrho \Pi + p \right) v^i
   \nonumber \\
   & & \qquad
       + \Pi^i_j v^j
       - {1 \over 4 \pi} \left[ \left( {\bf v} \times {\bf B} \right)
       \times {\bf B} \right]^i,
   \nonumber \\
   & & m^{ij} = \overline \varrho v^i v^j
       + \left( p +  {B^2 \over 8 \pi} \right) \delta^{ij}
       + \Pi^{ij}
       - {B^i B^j \over 4 \pi}.
   \label{D-def-NR}
\eea

In the spacetime metric, we have $\Psi = \Phi$, and Eq.\ (\ref{eq3-SRG})
gives $\chi_i = 0$. Although $\Psi$ does not affect the non-relativistic hydrodynamic
or MHD equations directly, the Newtonian gravity $\Phi$ naturally excites the
post-Newtonian potential $\Psi$ (Chandrasekhar 1965).

\section{General relativistic Electromagnetism}

The complete set of fully nonlinear and exact perturbation equations
with a general fluid component is presented in the Appendix of Hwang \&
Noh (2016). The presence of EM field can be accommodated in the formulation by
interpreting the contribution of the EM as fluid quantities with the Maxwell's
equations appended. The energy-momentum tensor of EM field is
\bea
   & & \widetilde T^{EM}_{ab}
       = {1 \over 4 \pi} \left( \widetilde F_{ac}
       \widetilde F_b^{\;\;c}
       - {1 \over 4} \widetilde g_{ab}
       \widetilde F_{cd} \widetilde F^{cd} \right).
   \label{Tab-EM}
\eea
The tildes
indicate
covariant quantities. The EM tensor can be decomposed as
\bea
   & & \widetilde F_{ab}
       \equiv \widetilde U_a \widetilde E_b
       - \widetilde U_b \widetilde E_a
       - \widetilde \eta_{abcd} \widetilde U^c \widetilde B^d,
   \label{Fab}
\eea
with
$\widetilde E_a \widetilde U^a \equiv 0 \equiv \widetilde B_a \widetilde U^a$;
$\widetilde U_a$ is a generic timelike four-vector normalized
so that
$\widetilde U^c \widetilde U_c \equiv -1$. With
\bea
   {}^*\widetilde F^{ab}
       \equiv {1 \over 2} \widetilde \eta^{abcd} \widetilde F_{cd}
       = \widetilde U^a \widetilde B^b
       - \widetilde U^b \widetilde B^a
       + \widetilde \eta^{abcd} \widetilde U_c \widetilde E_d,
   \label{*Fab}
\eea
we have
\bea
   & & \widetilde E_a = \widetilde F_{ab} \widetilde U^b, \quad
       \widetilde B_a
       = {}^*\widetilde F_{ab} \widetilde U^b.
\eea
Equation (\ref{Tab-EM}) can be written as
\bea
   & & \widetilde T^{EM}_{ab}
       = {1 \over 4 \pi} \bigg[
       \left( \widetilde E^2 + \widetilde B^2 \right)
       \left( \widetilde U_a \widetilde U_b
       + {1 \over 2} \widetilde g_{ab} \right)
       - \widetilde E_a \widetilde E_b
   \nonumber \\
   & & \qquad
       - \widetilde B_a \widetilde B_b
       + \left( \widetilde U_a \widetilde \eta_{bcde}
       + \widetilde U_b \widetilde \eta_{acde} \right)
       \widetilde E^c \widetilde B^d \widetilde U^e \bigg],
   \label{Tab-EM-cov}
\eea
and the fluid quantities in the $\widetilde U_a$ frame become
\bea
   & & \widetilde \mu_{EM}
       = 3 \widetilde p_{EM}
       = {1 \over 8 \pi} \left(
       \widetilde E^2 + \widetilde B^2 \right),
   \nonumber \\
   & &
       \widetilde q^{EM}_a
       = {1 \over 4 \pi} \widetilde \eta_{abcd}
       \widetilde E^b \widetilde B^c \widetilde U^d,
   \nonumber \\
   & &
       \widetilde \pi^{EM}_{ab}
       = - {1 \over 4 \pi}
       \left[ \widetilde E_a \widetilde E_b
       + \widetilde B_a \widetilde B_b
       - {1 \over 3} \widetilde h_{ab}
       \left( \widetilde E^2 + \widetilde B^2 \right) \right],
   \label{fluid-EM-cov}
\eea
with the fluid quantities in the $\widetilde U^a$-frame introduced as
\bea
   & & \widetilde T_{ab}
       = \widetilde \mu \widetilde U_a \widetilde U_b
       + \widetilde p \left( \widetilde g_{ab}
       + \widetilde U_a \widetilde U_b \right)
       + \widetilde q_a \widetilde U_b
       + \widetilde q_b \widetilde U_a
       + \widetilde \pi_{ab}.
   \nonumber \\
   \label{Tab-fluid-cov}
\eea
As we have non-vanishing $\widetilde q_a$ for EM field, in order to have
the nonlinear and exact perturbation formulation, we need to consider
$\widetilde q_a$ terms which are missing in our previous formulation.
In the ideal MHD considered in this work, the flux term vanishes for MHD,
see Eq.\ (\ref{fluid-MHD-cov}).

From
\bea
   & & {}^* \widetilde F^{ab}_{\;\;\;\; ;b} = 0, \quad
       \widetilde F^{ab}_{\;\;\;\; ;b}
       = {4 \pi \over c} \widetilde J_{ch}^a,
\eea
we have the four Maxwell's equations (Ellis 1973)
\bea
   & & \widetilde B^a_{\;\; ;b} \widetilde h^b_a
       = 2 \widetilde \omega^a \widetilde E_a,
   \label{Maxwell-cov-1} \\
   & & \widetilde h^a_b \widetilde B^b_{\;\; ;c} \widetilde U^c
       = \left( \widetilde \eta^a_{\;\;bcd} \widetilde U^d \widetilde \omega^c
       + \widetilde \sigma^a_{\;\;b}
       - {2 \over 3} \delta^a_b \widetilde \theta \right) \widetilde B^b
   \nonumber \\
   & & \qquad
       + \widetilde \eta^{abcd} \widetilde U_d
       \left( - \widetilde a_b \widetilde E_c
       + \widetilde E_{b;c} \right),
   \label{Maxwell-cov-2} \\
   & & \widetilde E^a_{\;\; ;b} \widetilde h^b_a
       = 4 \pi \widetilde \varrho_{ch}
       - 2 \widetilde \omega^a \widetilde B_a,
   \label{Maxwell-cov-3} \\
   & & \widetilde h^a_b \widetilde E^b_{\;\; ;c} \widetilde U^c
       = \left( \widetilde \eta^a_{\;\;bcd} \widetilde U^d \widetilde \omega^c
       + \widetilde \sigma^a_{\;\;b}
       - {2 \over 3} \delta^a_b \widetilde \theta \right)
       \widetilde E^b
   \nonumber \\
   & & \qquad
       + \widetilde \eta^{abcd} \widetilde U_d
       \left( \widetilde a_b \widetilde B_c
       - \widetilde B_{b;c} \right)
       - {4 \pi \over c} \widetilde j^a,
   \label{Maxwell-cov-4}
\eea
with $\widetilde h_{ab} \equiv \widetilde g_{ab} + \widetilde U_a \widetilde U_b$ the projection tensor. We
have decomposed the four-current as
\bea
   & & \widetilde J_{ch}^a
       \equiv \widetilde \varrho_{ch} c \widetilde U^a + \widetilde j^a, \quad
       \widetilde j_a \widetilde U^a \equiv 0,
   \label{current}
\eea
where the first and the second terms on the right-hand side
are the convection and conduction currents, respectively.
The covariant kinematic quantities $\widetilde \omega_a$, $\widetilde \sigma_{ab}$,
$\widetilde \theta$ and $\widetilde a_a$ are the vorticity vector, shear tensor,
expansion scalar, and acceleration vector, respectively, based on the generic
four-vector $\widetilde U_a$ (Ellis 1971, 1973).

For the fluid (comoving) frame four-vector we have $\widetilde U_a = \widetilde u_a$.
For the laboratory (normal) frame, we have $\widetilde U_a = \widetilde n_a$.
In the following, we set $\widetilde b_a \equiv \widetilde B^{(u)}_a$,
$\widetilde B_a \equiv \widetilde B^{(n)}_a$, and similarly for the electric field.

\section{General relativistic ideal MHD: derivation}
                                       \label{sec:FNLE-MHD-derivation}

The Ohm's law relates the conduction current in Eq.\ (\ref{current})
to the electric field in the comoving frame as (Jackson 1975)
\bea
   & & \widetilde j_a = \sigma \widetilde E_a^{(u)},
\eea
with $\sigma$ the electric conductivity. Ideal MHD
results from taking
perfectly conducting
limit (with $\sigma \rightarrow \infty ),$
so that
$\widetilde E_a^{(u)} = 0$, with non-vanishing $\widetilde j_a$.
In the following we consider
{\it ideal} MHD.

The ideal MHD equations may also be derived in the following
invariant (or coordinate-free) form:
\begin{equation}
\mathcal{L}_{\bar {\mathbf{U}}}\tilde {\mathbf{F}}=0,
\label{lieEvolution}
\end{equation}
which physically expresses the fact that the magnetic field lines are
frozen into the fluid and thus go with the flow. Here $\tilde {\mathbf{F}}$
is the Maxwell electromagnetic tensor regarded as a 2-form,
$\bar {\mathbf{U}}$ is the fluid 4-vector considered as
a vector field (and not as a covector field) as is sometimes denoted
by $\mathbf{U}^\#,$ and $\mathcal{L}$ denotes the Lie derivative.
Interestingly, the evolution equation [Eq.~(\ref{lieEvolution})]
thus expressed does not
involve the Riemannian (metric) structure of the manifold.
We derive Eq.~(\ref{lieEvolution}) as follows. The ideal
MHD assumption of infinite conductivity implies that in the
fluid rest frame $\mathbf{E}$ vanishes, or equivalently
$\tilde {\mathbf{F}}$ contracted with $\bar {\mathbf{U}}$
vanishes, which can be written as
$\mathbf{i}_{\bar {\mathbf{U}}} \tilde {\mathbf{F}}=0.$
Here $\mathbf{i}$ denotes the interior product.
The vanishing of the divergence of $\mathbf{B}$ and the Faraday
induction equation are expressed as
$\tilde {\mathbf{d}}
\tilde {\mathbf{F}}=0$
where $\tilde {\mathbf{d}}$
is the exterior derivative. Eq.~(\ref{lieEvolution})
follows by applying Cartan's magic formula (Abraham et al., 1988)
expressing the Lie derivative acting on a differential form
$\tilde {\boldsymbol{\alpha }}$ in the following way:
$\mathcal{L}_{\bar {\mathbf{X}}} \tilde {\boldsymbol{\alpha }}
=
(\mathbf{i}_{\bar {\mathbf{X}}} \circ \tilde {\mathbf{d}})
\tilde {\boldsymbol{\alpha }}
+
(\tilde {\mathbf{d}} \circ \mathbf{i}_{\bar {\mathbf{X}}})
\tilde {\boldsymbol{\alpha }}.$

The effect of the frozen in flux is to make the fluid behave much
like an anisotropic solid given that the flux lines cannot
move relative to the fluid. For our purposes the most important
consequence is the possibility of anisotropic stresses, which
cannot occur for an unmagnetized perfect fluid.

For ideal MHD Eqs.\ (\ref{Tab-EM-cov}) and (\ref{fluid-EM-cov}) become
\bea
   & & \widetilde T^{\rm MHD}_{ab}
       = {1 \over 4 \pi} \left[ \widetilde b^2 \left(
       \widetilde u_a \widetilde u_b
       + {1 \over 2} \widetilde g_{ab} \right)
       - \widetilde b_a \widetilde b_b \right],
   \\
   & & \widetilde \mu^{\rm MHD}
       = 3 \widetilde p^{\rm MHD}
       = {1 \over 8 \pi} \widetilde b^2, \quad
       \widetilde q^{\rm MHD}_a = 0,
   \nonumber \\
   & &
       \widetilde \pi^{\rm MHD}_{ab}
       = - {1 \over 4 \pi} \left[ \widetilde b_a
       \widetilde b_b
       - {1 \over 3} \widetilde b^2
       \left( \widetilde g_{ab} + \widetilde u_a \widetilde u_b \right) \right].
   \label{fluid-MHD-cov}
\eea

Now we should express $\widetilde b_a$ in terms of the magnetic field in the laboratory frame.

In order to express the fluid quantities in terms of
the
metric notation in Eq.\ (\ref{metric}),
the following quantities are useful. The exact inverse metric is (Hwang \& Noh 2013)
\bea
   & & \widetilde g^{00} = - {1 \over {\cal N}^2}, \quad
       \widetilde g^{0i}
       = - {\chi^i \over {\cal N}^2 (1 + 2 \varphi)},
   \nonumber \\
   & &
       \widetilde g^{ij} = {1 \over 1 + 2 \varphi}
       \left( \delta^{ij} - {\chi^i \chi^j \over {\cal N}^2 (1 + 2 \varphi) } \right),
\eea
with the index $0$ indicating $ct$; ${\cal N}$ is the lapse function
\bea
   & & {\cal N} = \sqrt{ 1 + 2 \alpha
       + {\chi^k \chi_k \over 1 + 2 \varphi}}.
\eea
The fluid four-vector becomes
\bea
   & &
       \widetilde u_i \equiv \gamma {v_i \over c}, \quad
       \widetilde u_0 = - \gamma \left( {\cal N}
       + {\chi^i \over 1 + 2 \varphi} {v_i \over c} \right),
   \nonumber \\
   & &
       \widetilde u^i
       = {\gamma \over 1 + 2 \varphi}
       \left( {v^i \over c} + {\chi^i \over {\cal N}} \right), \quad
       \widetilde u^0 = {1 \over {\cal N}} \gamma,
\eea
with the Lorentz factor
\bea
   & & \gamma
       \equiv {1 \over \sqrt{ 1 - {1 \over 1 + 2 \varphi}
       {v^2 \over c^2}}}.
   \label{Lorentz-factor}
\eea
The normal four-vector is
\bea
   & & \widetilde n_i \equiv 0, \quad
       \widetilde n_0 = - {\cal N}, \quad
       \widetilde n^i
       = {\chi^i \over {\cal N} (1 + 2 \varphi)}, \quad
       \widetilde n^0 = {1 \over {\cal N}}.
   \nonumber \\
\eea

For the field in the laboratory frame, using $\widetilde B_a \widetilde n^a = 0,$ we have
\bea
   & & \widetilde B_i \equiv B_i, \quad
       \widetilde B_0 = - {\chi_i B^i \over 1 + 2 \varphi}, \quad
       \widetilde B^i = {B^i \over 1 + 2 \varphi}, \quad
       \widetilde B^0 = 0,
   \nonumber \\
\eea
and similarly for the electric field $\widetilde E_a$ and the current density $\widetilde j_a$.
The index of $B_i$ is raised and lowered using $\delta_{ij}$ as the metric.

Using $0 = \widetilde E_{(u)}^a = \widetilde F^{ab} \widetilde u_b$,
and expressing $\widetilde F^{ab}$ in Eq.\ (\ref{Fab}) in the laboratory frame, we can show
\bea
   & & E^i = {- 1 \over \sqrt{1 + 2 \varphi}} \eta^{ijk}
       {v_j \over c} B_k
       = {- 1 \over \sqrt{1 + 2 \varphi}} {1 \over c}
       \left( {\bf v} \times {\bf B} \right)^i.
   \label{E-MHD-FNL}
\eea
It is useful to have
\bea
   & & \eta_{ijk}
       = - {\widetilde \eta_{0ijk} \over {\cal N} (1 + 2 \varphi)^{3/2}}, \quad
       \eta^{ijk} = {\cal N} (1 + 2 \varphi)^{3/2}
       \widetilde \eta^{0ijk},
   \nonumber \\
\eea
where indices of $\eta_{ijk}$ are raised and lowered using $\delta_{ij}$ as the metric.

Using $\widetilde b_a = {}^* \widetilde F_{ab} \widetilde u^b$
and expressing ${}^* \widetilde F_{ab}$ in Eq.\ (\ref{*Fab}) in the laboratory frame, we have
\bea
   & & \widetilde b_i = {1 \over \gamma} B_i
       + {\gamma \over 1 + 2 \varphi}
       {v_i \over c} B^j {v_j \over c},
   \nonumber \\
   & &
       \widetilde b_0 = - {1 \over 1 + 2 \varphi}
       \left( {\cal N} \gamma B^i {v_i \over c}
       + {\chi^i \over \gamma} B_i
       + {\gamma \chi^i \over 1 + 2 \varphi}
       {v_i \over c} B^j {v_j \over c}
       \right),
   \nonumber \\
   & &
       \widetilde b^i = {1 \over 1 + 2 \varphi} \left[
       {1 \over \gamma} B^i
       + {\gamma \over 1 + 2 \varphi}
       \left( {v^i \over c}
       + {1 \over {\cal N}} \chi^i \right)
       B^j {v_j \over c}
       \right],
   \nonumber \\
   & &
       \widetilde b^0 = {\gamma \over {\cal N} (1 + 2 \varphi)}
       B^i {v_i \over c}.
\eea
Using this, Eq.\ (\ref{fluid-MHD-cov}) gives
\bea
   & & \mu^{\rm MHD}
       = 3 p^{\rm MHD}
   \nonumber \\
   & & \qquad
       = {1 \over 8 \pi} {1 \over 1 + 2 \varphi} \left[
       {1 \over \gamma^2} B^2
       + {1 \over 1 + 2 \varphi} \left( B^i {v_i \over c} \right)^2 \right],
   \nonumber \\
   & & \Pi^{\rm MHD}_{ij}
       = {1 \over 4 \pi} \bigg\{
       - {1 \over \gamma^2} B_i B_j
   \nonumber \\
   & & \qquad
       + {1 \over 3} \delta_{ij} \left[
       {1 \over \gamma^2} B^2
        +{1 \over 1 + 2 \varphi} \left( B^k {v_k \over c} \right)^2 \right]
   \nonumber \\
   & & \qquad
       - {1 \over 1 + 2 \varphi}
       \left( B_i {v_j \over c} + B_j {v_i \over c} \right)
       B^k {v_k \over c}
   \nonumber \\
   & & \qquad
       + {1 \over 3} {1 \over 1 + 2 \varphi}
       \left[ B^2 - {2 \gamma^2\over 1 + 2 \varphi}
       \left( B^k {v_k \over c} \right)^2 \right]
       {v_i v_j \over c^2}
       \bigg\},
   \label{fluid-MHD}
\eea
where we set $\widetilde \mu \equiv \mu \equiv \varrho c^2$,
$\widetilde p \equiv p$ and $\widetilde \pi_{ij} \equiv \Pi_{ij}$
all in the fluid frame.
The indices of $\Pi_{ij}$ are raised and
lowered using $\delta_{ij}$ as the metric.
Using Eq.\ (\ref{E-MHD-FNL}) we have
\bea
   & & \mu^{\rm MHD}
       = 3 p^{\rm MHD}
       = {1 \over 8 \pi} {1 \over 1 + 2 \varphi} \left(
       B^2 - E^2 \right),
   \nonumber \\
   & & \Pi^{\rm MHD}_{ij}
       = {1 \over 4 \pi} \bigg[
       - E_i E_j - B_i B_j
       + {1 \over 3} \delta_{ij} \left(
       2 E^2 + B^2 \right)
   \nonumber \\
   & & \qquad
       + {2 \over 3} {1 \over 1 + 2 \varphi}
       \left( E^2 - B^2 \right) \gamma^2
       {v_i v_j \over c^2}
       \bigg].
   \label{fluid-MHD-E}
\eea
These relations expressing the fluid quantities
in the comoving (energy) frame in terms of the fields in the
laboratory (normal) frame, appear asymmetric in the fields.
This is because although the energy-momentum tensor in Eq.\ (\ref{Tab-fluid-cov})
is frame invariant, the fluid quantities in Eq.\ (\ref{fluid-EM-cov}) are not.

Using the fluid quantities in Eq.\ (\ref{fluid-MHD}),
by replacing $\mu \rightarrow \mu + \mu^{\rm MHD}$ etc.,
the fully nonlinear and exact perturbation equations in Hwang \& Noh (2016) are
now complete in the presence of MHD. A complete set of equations is presented in the Appendix.

In the presence of EM field, we also need to include the Maxwell equations.
Taking the laboratory frame Eqs.\ (\ref{Maxwell-cov-1})-(\ref{Maxwell-cov-4})
gives Eqs.\ (\ref{Maxwell-1-FNL})-(\ref{Maxwell-4-FNL}).

Equations in this section and in the Appendix are fully general in Einstein's gravity,
under the conditions stated below Eq.\ (\ref{metric}), with MHD.
We have not yet imposed temporal gauge condition.

\section{Weak gravity limit: derivation}
                                        \label{sec:SR-MHD-derivation}

Now, using the fully nonlinear and exact formulation of GR MHD presented in the Appendix, we
prove equations in Sec.\ \ref{sec:SRG} by taking the weak gravity and action-at-a-distance
limit in Eq.\ (\ref{WG-conditions}).

The ADM momentum constraint equation in Eq.\ (\ref{eq3}) becomes
\bea
   & & {2 \over 3} \kappa_{,i}
       + {c \over {\cal N}}
       \left( {2 \over 3} \Delta \chi_{,i}
       + {1 \over 2} \Delta \chi^{(v)}_i \right)
       = - {8 \pi G \over c^4} \bigg[
       \left( \mu + p \right) \gamma^2 v_i
   \nonumber \\
   & & \qquad
       + \Pi_{ij} v^j
       + {1 \over 4 \pi} \left( B^2 v_i - B_i B^j v_j \right)
       \bigg]
       = - {8 \pi G \over c^2} m_i,
   \label{eq3-SR}
\eea
where we decomposed $\chi_i \equiv \chi_{,i} + \chi_i^{(v)}$.

Now we {\it take} the maximal slicing as the temporal gauge condition
\bea
   & & \kappa \equiv 0.
\eea
Thus
\bea
   & & \chi = - {12 \pi G \over c^3} \Delta^{-2} \nabla^i m_i,
   \label{eq3-SRG-scalar} \\
   & & \chi^{(v)}_i = - {16 \pi G \over c^3} \Delta^{-1}
       \left( m_i
       - \Delta^{-1} \nabla_i \nabla^j m_j \right).
   \label{eq3-SRG-vector}
\eea
These give Eq.\ (\ref{eq3-SRG}).

Considering Eq.\ (\ref{relativistic-order}), Eqs.\ (\ref{eq3-SRG-scalar})
and (\ref{eq3-SRG-vector}) gives
\bea
   & & \chi_{,i} \sim \chi^{(v)}_i
       \sim \gamma^2 {t_\ell^2 \over t_g^2} {v_i \over c},
\eea
thus we have
\bea
   & & \chi_{,i} \sim \chi^{(v)}_i \ll {v_i \over c},
   \label{chi-order}
\eea
and ${\cal N} = 1$.

The energy and momentum conservation equations in Eqs.\ (\ref{eq6}) and (\ref{eq7}), respectively, give
\bea
   & & \dot E
       + \nabla_i m^i c^2
       = - \overline \varrho \left( 2 \Phi - \Psi \right)_{,i} v^i,
   \label{eq6-SRG}
   \\
   & & \dot m^i + \nabla_j m^{ij}
       = - \overline \varrho \Phi^{,i}.
\eea
The continuity equation in Eq.\ (\ref{eq0}) gives
\bea
   & & \dot D
       + \nabla_i \left( D v^i \right)
       = 0.
   \label{eq0-SRG}
\eea
These are three equations in Eq.\ (\ref{SR-MHD-eqs}). Derivation of Eq.\ (\ref{eq6-SRG})
deserves a special comment. It is important to carefully keep the gravity term in right-hand-side
as explained above Eq.\ (\ref{E-conservation-NR}). As we examine Eq.\ (\ref{eq6}) we notice that
the first term in the equation leads to $3 \overline \varrho \dot \Psi$ to the gravity part
which is of the same order as we consider $\dot \Psi \sim \Psi_{,i} v^i$. This term,
however, exactly cancels the $\chi^i$-term in the second line because of Eq.\ (\ref{k-def}).

The trace of ADM propagation and energy constraint equations in Eqs.\ (\ref{eq4}) and (\ref{eq2}),
respectively, give Eqs.\ (\ref{eq4-SRG}) and (\ref{eq2-SRG}). We can show that the traceless part
of ADM propagation in Eq.\ (\ref{eq5}) simply gives a combination of Eqs.\ (\ref{eq4-SRG}) and (\ref{eq2-SRG}).

Finally, Eq.\ (\ref{eq1}) gives
\bea
   & & c \Delta \chi = 3 {\dot \Psi \over c^2},
   \label{k-def}
\eea
and using Eqs.\ (\ref{eq2-SRG}), (\ref{eq3-SRG-scalar}), and (\ref{eq6-SRG}),
we can show that this is naturally valid. This calls for
comment as the validity of Eq.\ (\ref{k-def}) in our approximation misses
the gravity term in Eq.\ (\ref{eq6-SRG}) in the derivation. This is because Eq.\ (\ref{k-def})
is already a first-order post-Newtonian (1PN) order whereas our
approximation is zeroth-order PN (0PN) in gravity while exact in matter part.
Equation (\ref{eq1}) is the definition of the trace of extrinsic curvature
$K^i_i$ and its PN nature is presented in Eq.\ (55) of Hwang et al (2008).

Thus using the complete set of Einstein's equations,
we have shown the consistency of our SR MHD equations with weak self-gravity presented in Sec.\ \ref{sec:SRG}.

In the weak gravity limit,
the effect of gravity does not appear
in the Maxwell equations. Equations
(\ref{Maxwell-1-FNL})-(\ref{Maxwell-4-FNL}) become
\bea
   & & B^i_{\;\;,i} = 0,
   \label{Maxwell-1-SRG} \\
   & & \dot B^i
       = \left( v^i B^j - v^j B^i \right)_{,j}
       = \left[ \nabla \times \left( {\bf v} \times {\bf B} \right)
       \right]^i,
   \label{Maxwell-2-SRG} \\
   & & \nabla \cdot {\bf E} = 4 \pi \varrho_{ch},
   \label{Maxwell-3-SRG} \\
   & & {1 \over c} \dot {\bf E}
       = \nabla \times {\bf B} - {4 \pi \over c} {\bf j},
   \label{Maxwell-4-SRG}
\eea
with
\bea
   & & {\bf E}
       = - {1 \over c}
       {\bf v} \times {\bf B},
   \label{E-MHD}
\eea
in the ideal MHD. Equations (\ref{Maxwell-1-SRG}) and (\ref{Maxwell-2-SRG}) can be written as
\bea
   & & \nabla \cdot {\bf B} = 0, \quad
       {1 \over c} \dot {\bf B}
       = - \nabla \times {\bf E}.
\eea
These are the well known form of Maxwell's equations valid for SR MHD with Eq.\ (\ref{E-MHD}).

Thus we have derived the equations in Sec.\ \ref{sec:SRG}.

\section{Discussion}

The two formulations of relativistic MHD are new results in this work.
These are the GR MHD in the fully nonlinear
and exact perturbation formulation of Einstein's gravity, and the SR MHD with weak gravity.

The fully nonlinear and exact perturbation formulation of ideal MHD in Einstein's gravity is derived in Sec.\
\ref{sec:FNLE-MHD-derivation} and the equations are presented in the Appendix. These are exact equations using
perturbation variables imposed on the Minkowski metric, see Eq.\ (\ref{metric}). We have ignored the
transverse-traceless perturbation, but including this as well as not imposing the spatial gauge condition
can be trivially achieved. However, the equations may look complicated though.
For a general hydrodynamic fluid, see Gong et al (2017).

By taking the weak gravity and action-at-a-distance limits, we derived a consistent formulation of fully SR MHD
with weak gravity, see Secs.\ \ref{sec:SRG} and \ref{sec:SR-MHD-derivation}. We show that the role of gravity
on the dynamics is effectively the same as in the Newtonian limit. However, the SR effects of the fluid and EM
field affects the metric, thus gravitational potentials, and these could affect the gravitational lensing, see
Sec.\ \ref{sec:SRG-lensing}. The SR effects, if important, might cause overestimation of the lensing mass, see
Eq.\ (\ref{lensing-potential}).

The weak gravity formulation is derived in the maximal slicing ($\kappa \equiv 0$) gauge,
which is
the unique gauge choice
with a consistent weak gravity limit.
A similar choice of the zero-shear gauge (often termed as longitudinal or
conformal Newtonian gauge), taking $\chi \equiv 0$ as the slicing condition, leads to an inconsistent result
by
omitting
the pressure term (see Sec.\ 2.3 in Hwang \& Noh 2016).

Our weak gravity approximation is complementary to the PN approximation. The PN approximation perturbatively
expands both gravity and matter consistently. To 1PN order we keep $\Phi/c^2 \sim \Psi/c^2 \sim v^2/c^2 \sim
\Pi/c^2 \sim p/ (\overline \varrho c^2) \sim \Pi_{ij}/ (\overline \varrho c^2) \sim B^2/ (\overline \varrho
c^2)$, etc (Chandrasekhar 1965; Greenberg 1971; Hwang et al 2008). In this sense our weak gravity
approximation with full SR is a 0PN approximation in gravity ($\Phi$, $\Psi$ and $\chi_i$) but exact in the
matter part (internal energy, pressure, stress, magnetic field, etc), and thus handles the matter part to
$\infty$PN order. It is not {\it a priori} clear that such an asymmetric formulation is possible,
but here we have shown that it is indeed possible.
Extending the program to include gravity to 1PN order might be feasible.

The validity of our approximation can be checked by
comparing with a full numerical relativity simulation in the same gauge
(the maximal slicing together with our spatial gauge condition taken).

\vskip .5cm
%
%

We wish to thank Professor Dongsu Ryu for encouraging us to pursue this subject. H.N.\ was supported by National Research Foundation of Korea funded by the Korean Government (No.\ 2018R1A2B6002466).
J.H.\ was supported by Basic Science Research Program through the National Research Foundation (NRF) of Korea funded by the Ministry of Science, ICT and future Planning (No.\ 2016R1A2B4007964 and No.\ 2018R1A6A1A06024970).
MB thanks SKA South Africa as well as an NRF KIC grant for partial support.

\begin{widetext}
\section*{Appendix: General relativistic ideal MHD equations}

Here we present the complete set of Einstein's equation
for a general fluid
with
ideal MHD.
These equations without MHD are presented in the Appendix B of Hwang \& Noh (2016). The
MHD parts can be included by replacing the fluid quantities as
$\mu \rightarrow \mu + \mu^{\rm MHD}$, $p \rightarrow p + p^{\rm MHD}$,
$\Pi_{ij} \rightarrow \Pi_{ij} + \Pi_{ij}^{\rm MHD}$ using Eq.\ (\ref{fluid-MHD}).
The equations are generally valid irrespective of
the temporal gauge choice (slicing, hypersurface).
All spatial indices in the Appendix are raised and
lowered using $\delta_{ij}$ as the metric.

\noindent
The definition of $\kappa$ (the trace of extrinsic curvature $K^i_i \equiv \kappa/c$):
\bea
   & & \kappa
       \equiv
       - {1 \over {\cal N} (1 + 2 \varphi)}
       \left[ 3 \dot \varphi
       + c \left( \chi^k_{\;\;,k}
       + {\chi^{k} \varphi_{,k} \over 1 + 2 \varphi} \right)
       \right].
   \label{eq1}
\eea
ADM energy constraint:
\bea
   & & {4 \pi G \over c^2} \mu
       + {c^2 \Delta \varphi \over (1 + 2 \varphi)^2}
       = {1 \over 6} \kappa^2
       + {3 \over 2} {c^2 \varphi^{,i} \varphi_{,i} \over (1 + 2 \varphi)^3}
       - {c^2 \over 4} \overline{K}^i_j \overline{K}^j_i
   \nonumber \\
   & & \qquad
       - {4 \pi G \over c^2} \left\{
       \left( \mu + p \right)
       \left( \gamma^2 - 1 \right)
       + {1 \over 1 + 2 \varphi} \Pi^i_i
       + {1 \over 8 \pi} {1 \over 1 + 2 \varphi} \left[
       \left( 2 - {1 \over \gamma^2} \right) B^2
       - {1 \over 1 + 2 \varphi} \left( B^i {v_i \over c} \right)^2 \right] \right\}.
   \label{eq2}
\eea
ADM momentum constraint:
\bea
   & & {2 \over 3} \kappa_{,i}
       + {c \over {\cal N} ( 1 + 2 \varphi )}
       \left( {1 \over 2} \Delta \chi_i
       + {1 \over 6} \chi^k_{\;\;,ki} \right)
       =
       {c \over {\cal N} ( 1 + 2 \varphi)}
       \bigg\{
       \left( {{\cal N}_{,j} \over {\cal N}}
       - {\varphi_{,j} \over 1 + 2 \varphi} \right)
       \left[ {1 \over 2} \left( \chi^{j}_{\;\;,i} + \chi_i^{\;,j} \right)
       - {1 \over 3} \delta^j_i \chi^k_{\;\;,k} \right]
   \nonumber \\
   & & \qquad \qquad
       - {\varphi^{,j} \over (1 + 2 \varphi)^2}
       \left( \chi_{i} \varphi_{,j}
       + {1 \over 3} \chi_{j} \varphi_{,i} \right)
       + {{\cal N} \over 1 + 2 \varphi} \nabla_j
       \left[ {1 \over {\cal N}} \left(
       \chi^{j} \varphi_{,i}
       + \chi_{i} \varphi^{,j}
       - {2 \over 3} \delta^j_i \chi^{k} \varphi_{,k} \right) \right]
       \bigg\}
   \nonumber \\
   & & \qquad \qquad
       - {8 \pi G \over c^4} \left[
       \left( \mu + p \right)
       \gamma^2 v_i
       + {1 \over 1 + 2 \varphi} \Pi_{ij} v^j
        + {1 \over 4 \pi} {1 \over 1 + 2 \varphi} \left( B^2 {v_i} - B_i B^j {v_j} \right) \right].
   \label{eq3}
\eea
Trace of ADM propagation:
\bea
   & & - {4 \pi G \over c^2} \left( \mu + 3 p \right)
       + {1 \over {\cal N}} \dot \kappa
       + {c^2 \Delta {\cal N} \over {\cal N} (1 + 2 \varphi)}
       = {1 \over 3} \kappa^2
       - {c \over {\cal N} (1 + 2 \varphi)} \left(
       \chi^{i} \kappa_{,i}
       + c {\varphi^{,i} {\cal N}_{,i} \over 1 + 2 \varphi} \right)
       + c^2 \overline{K}^i_j \overline{K}^j_i
   \nonumber \\
   & & \qquad
       + {8 \pi G \over c^2} \left\{
       \left( \mu + p \right)
       \left( \gamma^2 - 1 \right)
       + {1 \over 1 + 2 \varphi} \Pi^i_i
       + {1 \over 8 \pi} {1 \over 1 + 2 \varphi} \left[
       \left( 2 - {1 \over \gamma^2} \right) B^2
       - {1 \over 1 + 2 \varphi} \left( B^i {v_i \over c} \right)^2 \right] \right\}.
   \label{eq4}
\eea
Tracefree ADM propagation:
\bea
   & & \left( {1 \over {\cal N}} {\partial \over \partial t}
       - \kappa
       + {c \chi^{k} \over {\cal N} (1 + 2 \varphi)} \nabla_k \right)
       \bigg\{ {c \over {\cal N} (1 + 2 \varphi)}
       \left[
       {1 \over 2} \left( \chi^i_{\;\;,j} + \chi_j^{\;,i} \right)
       - {1 \over 3} \delta^i_j \chi^k_{\;\;,k}
       - {1 \over 1 + 2 \varphi} \left( \chi^{i} \varphi_{,j}
       + \chi_{j} \varphi^{,i}
       - {2 \over 3} \delta^i_j \chi^{k} \varphi_{,k} \right)
       \right] \bigg\}
   \nonumber \\
   & & \qquad
       - {c^2 \over ( 1 + 2 \varphi)}
       \left[ {1 \over 1 + 2 \varphi}
       \left( \nabla^i \nabla_j - {1 \over 3} \delta^i_j \Delta \right) \varphi
       + {1 \over {\cal N}}
       \left( \nabla^i \nabla_j - {1 \over 3} \delta^i_j \Delta \right) {\cal N} \right]
       =
       {c^2 \over {\cal N}^2 (1 + 2 \varphi)^2}
       \bigg[
       {1 \over 2} \left( \chi^{i,k} \chi_{j,k}
       - \chi^{k,i} \chi_{k,j} \right)
   \nonumber \\
   & & \qquad
       + {1 \over 1 + 2 \varphi} \left(
       \chi^{k,i} \chi_k \varphi_{,j}
       - \chi^{i,k} \chi_j \varphi_{,k}
       + \chi_{k,j} \chi^k \varphi^{,i}
       - \chi_{j,k} \chi^i \varphi^{,k} \right)
       + {2 \over (1 + 2 \varphi)^2} \left(
       \chi^{i} \chi_{j} \varphi^{,k} \varphi_{,k}
       - \chi^{k} \chi_{k} \varphi^{,i} \varphi_{,j} \right) \bigg]
   \nonumber \\
   & & \qquad
       - {c^2 \over (1 + 2 \varphi)^2}
       \left[ {3 \over 1 + 2 \varphi}
       \left( \varphi^{,i} \varphi_{,j}
       - {1 \over 3} \delta^i_j \varphi^{,k} \varphi_{,k} \right)
       + {1 \over {\cal N}} \left(
       \varphi^{,i} {\cal N}_{,j}
       + \varphi_{,j} {\cal N}^{,i}
       - {2 \over 3} \delta^i_j \varphi^{,k} {\cal N}_{,k} \right) \right]
   \nonumber \\
   & & \qquad
       + {8 \pi G \over c^2} \Bigg\{
       \left( \mu +  p \right)
       \left[ {\gamma^2 v^i v_j \over c^2 (1 + 2 \varphi)}
       - {1 \over 3} \delta^i_j \left( \gamma^2 - 1 \right)
       \right]
       + {1 \over 1 + 2 \varphi} \left( \Pi^i_j
       - {1 \over 3} \delta^i_j \Pi^k_k \right)
       + {1 \over 4 \pi} {1 \over 1 + 2 \varphi} \bigg\{
       - {1 \over \gamma^2} B^i B_j
   \nonumber \\
   & & \qquad
       - {1 \over 1 + 2 \varphi}
       \left( B^i {v_j \over c} + B_j {v^i \over c} \right)
       B^k {v_k \over c}
       + {1 \over 1 + 2 \varphi} B^2 {v^i v_j \over c^2}
       - {1 \over 3} \delta^i_j \left[
       \left( 1 - {2 \over \gamma^2} \right) B^2
       - {2 \over 1 + 2 \varphi} \left( B^k {v_k \over c} \right)^2 \right] \bigg\} \Bigg\}.
   \label{eq5}
\eea
ADM energy conservation:
\bea
   & &
       {1 \over c} \Bigg\{
       \left( 1 + 2 \varphi \right)^{3/2}
       \Bigg[ \mu + \left( \mu + p \right) \left( \gamma^2 - 1 \right)
       + {1 \over 1 + 2 \varphi} \Pi^i_i
       + {1 \over 1 + 2 \varphi}
       {1 \over 8 \pi}
       \left[ \left( 2 - {1 \over \gamma^2} \right) B^2
       - {1 \over 1 + 2 \varphi}
       \left( B^i {v_i \over c} \right)^2 \right]
       \Bigg]
       \Bigg\}^{\displaystyle\cdot}
   \nonumber \\
   & & \qquad
       + \Bigg\{
       \left( 1 + 2 \varphi \right)^{1/2} {\cal N} \left[
       \left( \mu + p \right) \gamma^2 {v^i \over c}
       + {1 \over 1 + 2 \varphi} \Pi^i_j {v^j \over c}
       + {1 \over 1 + 2 \varphi} {1 \over 4 \pi}
       \left( B^2 {v^i \over c}
       - B^i B^j {v_j \over c} \right) \right]
       + \left( 1 + 2 \varphi \right)^{1/2}
       \chi^i
   \nonumber \\
   & & \qquad
       \times
       \left\{ \mu + \left( \mu + p \right)
       \left( \gamma^2 - 1 \right)
       + {1 \over 1 + 2 \varphi} \Pi^j_j
       + {1 \over 1 + 2 \varphi} {1 \over 8 \pi}
       \left[ \left( 2 - {1 \over \gamma^2} \right) B^2
       - {1 \over 1 + 2 \varphi}
       \left( B^j {v_j \over c} \right)^2 \right]
       \right\}
       \Bigg\}_{,i}
   \nonumber \\
   & & \qquad
       = - \left( 1 + 2 \varphi \right)^{-1/2}
       \left[ {1 \over c} \dot \varphi \delta^{ij}
       + \chi^{i,j}
       - {1 \over 1 + 2 \varphi}
       \left( 2 \chi^i \varphi^{,j}
       - \delta^{ij} \chi^k \varphi_{,k} \right) \right]
       \Bigg\{ \left( \mu + p \right) \gamma^2 {v_i v_j \over c^2}
       + \left( 1 + 2 \varphi \right) p \delta_{ij}
       + \Pi_{ij}
   \nonumber \\
   & & \qquad
       + {1 \over 4 \pi} \left\{ - {1 \over \gamma^2} B_i B_j
       - {1 \over 1 + 2 \varphi}
       \left( B_i {v_j \over c} + B_j {v_i \over c} \right)
       B^k {v_k \over c}
       + {1 \over 1 + 2 \varphi} B^2 {v_i v_j \over c^2}
       + {1 \over 2} \delta_{ij}
       \left[ {1 \over \gamma^2} B^2
       + {1 \over 1 + 2 \varphi}
       \left( B^k {v_k \over c} \right)^2 \right] \right\}
       \Bigg\}
   \nonumber \\
   & & \qquad
       - \left( 1 + 2 \varphi \right)^{1/2} {\cal N}_{,i}
       \left[ \left( \mu + p \right) \gamma^2 {v^i \over c}
       + {1 \over 1 + 2 \varphi} \Pi^i_j {v^j \over c}
       + {1 \over 1 + 2 \varphi} {1 \over 4 \pi}
       \left( B^2 {v^i \over c}
       - B^i B^j {v_j \over c} \right) \right].
   \label{eq6}
\eea
ADM momentum conservation:
\bea
   & & {1 \over c} \left\{
       \left( 1 + 2 \varphi \right)^{3/2}
       \left[ \left( \mu + p \right) \gamma^2 {v_i \over c}
       + {1 \over 1 + 2 \varphi} \Pi_i^j {v_j \over c}
       + {1 \over 1 + 2 \varphi}
       {1 \over 4 \pi}
       \left( B^2 {v_i \over c}
       - B_i B^j {v_j \over c} \right) \right]
       \right\}^{\displaystyle\cdot}
   \nonumber \\
   & & \qquad
       + \Bigg\{
       \left( 1 + 2 \varphi \right)^{1/2}
       {\cal N} \Bigg[ \left( \mu + p \right)
       \gamma^2 {v_i v^j \over c^2}
       + \left( 1 + 2 \varphi \right) p \delta_i^j
       + \Pi_i^j
   \nonumber \\
   & & \qquad
       + {1 \over 4 \pi}
       \left\{
       - {1 \over \gamma^2} B_i B^j
       - {1 \over 1 + 2 \varphi}
       \left( B_i {v^j \over c} + B^j {v_i \over c} \right)
       B^k {v_k \over c}
       + {1 \over 1 + 2 \varphi} B^2 {v_i v^j \over c^2}
       + {1 \over 2} \delta_i^j
       \left[ {1 \over \gamma^2} B^2
       + {1 \over 1 + 2 \varphi} \left( B^k {v_k \over c} \right)^2
       \right] \right\} \Bigg]
   \nonumber \\
   & & \qquad
       + \left( 1 + 2 \varphi \right)^{1/2} \chi^j \left[
       \left( \mu + p \right) \gamma^2 {v_i \over c}
       + {1 \over 1 + 2 \varphi} \Pi_i^k {v_k \over c}
       + {1 \over 1 + 2 \varphi}
       {1 \over 4 \pi}
       \left( B^2 {v_i \over c} - B_i B^k {v_k \over c} \right)
       \right]
       \Bigg\}_{,j}
   \nonumber \\
   & & \qquad
       = \left( 1 + 2 \varphi \right)^{3/2} \Bigg\{
       {{\cal N} \varphi_{,i} \over 1 + 2 \varphi}
       \left[ \left( \mu + p \right) \left( \gamma^2 - 1 \right)
       + 3 p + {1 \over 1 + 2 \varphi} \Pi^j_j \right]
       - {\cal N}_{,i}
       \left[ \mu + \left( \mu + p \right) \left( \gamma^2 - 1 \right)
       + {1 \over 1 + 2 \varphi} \Pi^j_j \right]
   \nonumber \\
   & & \qquad
       + \left( {{\cal N} \varphi_{,i} \over 1 + 2 \varphi}
       - {\cal N}_{,i} \right)
       {1 \over 1 + 2 \varphi } {1 \over 4 \pi}
       \left[ B^2 \left( 1 - {1 \over 2 \gamma^2} \right)
       - {1 \over 2} {1 \over 1 + 2 \varphi}
       \left( B^k {v_k \over c} \right)^2 \right]
   \nonumber \\
   & & \qquad
       - \left( {\chi^j \over 1 + 2 \varphi} \right)_{,i}
       \left[ \left( \mu + p \right) \gamma^2 {v_j \over c}
       + {1 \over 1 + 2 \varphi} \Pi_j^k {v_k \over c}
       + {1 \over 1 + 2 \varphi } {1 \over 4 \pi}
       \left( B^2 {v_j \over c} -B_j B^k {v_k \over c} \right) \right]
       \Bigg\},
   \label{eq7}
\eea
Continuity equation, $(\overline \varrho u^c)_{;c} = 0$:
\bea
   & & \left[ {\partial \over \partial t}
       + {1 \over 1 + 2 \varphi}
       \left( {\cal N} v^i + c \chi^i \right)
       \nabla_i
       - {\cal N} \kappa
       + { ( {\cal N} v^i )_{,i} \over 1 + 2 \varphi}
       +{ {\cal N} v^i \varphi_{,i} \over ( 1 + 2 \varphi )^2}
       \right] \overline \varrho \gamma
       = 0.
   \label{eq0}
\eea
Maxwell's equations:
\bea
   & & \nabla \cdot \left( \sqrt{ 1 + 2 \varphi } {\bf B} \right)
       = 0,
   \label{Maxwell-1-FNL} \\
   & & \left( \sqrt{ 1 + 2 \varphi } {\bf B}
       \right)^{\displaystyle\cdot}
       = \nabla \times \left[ {1 \over \sqrt{1 + 2 \varphi}}
       \left( {\cal N} {\bf v} + c \vec{\chi} \right)
       \times {\bf B} \right],
   \label{Maxwell-2-FNL} \\
   & & - {1 \over (1 + 2 \varphi)^{3/2}}
       {1 \over c} \nabla \cdot
       \left( {\bf v} \times {\bf B} \right)
       = 4 \pi \varrho_{ch},
   \label{Maxwell-3-FNL} \\
   & & - {1 \over c^2} \left( {\bf v} \times {\bf B}
       \right)^{\displaystyle{\cdot}}
       = \nabla \times \left[ {\cal N} {\bf B}
       - {1 \over c} {1 \over 1 + 2 \varphi}
       \left( {\bf v} {\bf B} \cdot \vec{\chi}
       - {\bf v} \cdot \vec{\chi} {\bf B} \right) \right]
       - 4 \pi \sqrt{1 + 2 \varphi} \left( \varrho_{ch} \vec{\chi}
       + {1 \over c} {\cal N} \vec{j} \right).
   \label{Maxwell-4-FNL}
\eea
We have
\bea
   & & \overline{K}^i_j \overline{K}^j_i
       = {1 \over {\cal N}^2 (1 + 2 \varphi)^2}
       \bigg\{
       {1 \over 2} \chi^{i,j} \left( \chi_{i,j} + \chi_{j,i} \right)
       - {1 \over 3} \chi^i_{\;\;,i} \chi^j_{\;\;,j}
       - {4 \over 1 + 2 \varphi} \left[
       {1 \over 2} \chi^i \varphi^{,j} \left(
       \chi_{i,j} + \chi_{j,i} \right)
       - {1 \over 3} \chi^i_{\;\;,i} \chi^j \varphi_{,j} \right]
   \nonumber \\
   & &
       + {2 \over (1 + 2 \varphi)^2} \left(
       \chi^{i} \chi_{i} \varphi^{,j} \varphi_{,j}
       + {1 \over 3} \chi^i \chi^j \varphi_{,i} \varphi_{,j} \right) \bigg\}, \quad
       \Pi^i_i = {1 \over 1 + 2 \varphi} \Pi_{ij} {v^i v^j \over c^2}.
   \label{K-bar-eq}
\eea
\end{widetext}

One of the following conditions can be imposed as the temporal gauge condition:
(1) maximal slicing (i.e., $\kappa \equiv 0$); (2) zero-shear gauge
(i.e., setting the longitudinal part of $\chi_i$ to zero, so that $\chi \equiv 0$);
or (3) comoving gauge (i.e., setting the longitudinal part of $v_i$ to zero).
These three gauge conditions, as well as various linear combinations thereof, completely remove
both the spatial and temporal gauge modes.
Another possible gauge condition is synchronous gauge, setting $\alpha \equiv 0,$
but this gauge choice fails to fix the gauge modes completely. [For a discussion
of gauge transformation at linear order, see Eq.\ (79) in Hwang \& Noh (2016), and for higher
(nonlinear) orders, see Noh \& Hwang (2004).]

%
%


\end{document}